\documentstyle[aps,psfig]{revtex}

\newcommand{\ddd}{{\mathrm d}}
\newcommand{\Hconf}{{\mathcal H}}

\newcommand{\ETC}{etc}

\newcommand{\IE}{i.e.}
\newcommand{\EG}{e.g.}
\newcommand{\ETAL}{{\it et al.}}

\newcommand{\EQ}[1]{eq.~(#1)}

\newcommand{\EQNS}[1]{eqns.~(#1)}
\newcommand{\EEQNS}[1]{Eqns.~(#1)}
\newcommand{\FIG}[1]{Fig.~{#1}}
\newcommand{\FFIG}[1]{Fig.~{#1}}
\newcommand{\FIGS}[1]{Figs.~{#1}}

\newcommand{\XK}{k_c}
\newcommand{\XH}{\Gamma}
\newcommand{\XM}{\dot\mu }

\newcommand{\pbp}{\Psi}
\newcommand{\pbq}{\Phi}

\begin{document}

\title{Topological Defects and CMB anisotropies~: Are the
predictions reliable~?}

\author{
Alain Riazuelo$^{\hbox{\scriptsize a}}$,
Nathalie Deruelle$^{\hbox{\scriptsize a,b,c}}$
and Patrick Peter$^{\hbox{\scriptsize a,d}}$
}

\address{~\\
$^{\hbox{\scriptsize a}}$ D\'epartement d'Astrophysique
Relativiste et de Cosmologie, \\ UMR 8629 du CNRS, Observatoire de
Paris, 5 place J. Janssen, 92195 Meudon,
France \\~\\
$^{\hbox{\scriptsize b}}$ Department of Applied Mathematics and
Theoretical Physics, \\ University of Cambridge, Silver Street,
Cambridge, CB3 9EW, England \\~\\
$^{\hbox{\scriptsize c}}$ Institut des Hautes \'Etudes
Scientifiques, \\ 91140 Bures-sur-Yvette, France \\~\\
$^{\hbox{\scriptsize d}}$ Institut d'Astrophysique de Paris,
UPR 341 du CNRS, \\ 98 bis boulevard Arago, 75014 Paris \\~}

\date{15 december 1999}

\maketitle

\begin{abstract}
We consider a network of topological defects which can partly decay
into neutrinos, photons, baryons, or Cold Dark Matter. We find that
the degree-scale amplitude of the cosmic microwave background (CMB)
anisotropies as well as the shape of the matter power spectrum can be
considerably modified when such a decay is taken into account. We
conclude that present predictions concerning structure formation by
defects might be unreliable.
\end{abstract}


\section{Introduction}
\label{sec_intro}

Two challenging paradigms to explain structure formation in the
Universe are currently developped, namely cosmological
inflation~\cite{inflation} and topological defects~\cite{td}. On the
one hand, inflation is a simple theory, based on the linear evolution
of acausal and coherent initial perturbations produced during an
accelerated phase of expansion of the early Universe. Several public
codes are available, notably CMBFAST~\cite{cmbcode}, to compute the
CMB anisotropies of a given model in few minutes, and these
predictions are robust. On the other hand, topological defects, which
are supposed to have formed after a cosmological phase transition in
the early Universe, are much more difficult to handle, because of the
highly non-linear structure of their dynamics.  After some pioneering
work by Bennett, Bouchet, Stebbins~\cite{bbs}, Shellard,
Allen~\cite{early_sa}, Perivolaropoulos~\cite{perivar}, Caldwell
\ETAL~\cite{caldwell}, the first detailed predictions of some defect
models have been published only recently. In particular, Turok, then
Pen, Seljak and Turok~\cite{turok} have considered global defects, as
well as Durrer and collaborators~\cite{durrer} who studied more
carefully the large $N$ limit. Following Vincent, Hindmarsh and
Sakellariadou~\cite{vincent}, Battye, Albrecht and
Robinson~\cite{battye}, have studied a network of line-like segments
with given correlation length and velocity distribution, as well as
Pogosian and Vachaspati~\cite{vachas} who also considered the wiggly
structure of the strings. Allen, Shellard and
collaborators~\cite{shellard}, and Contaldi, Hindmarsh and
Magueijo~\cite{magueijo} have considered local cosmic strings.

Several ground experiments (Saskatoon, Python V, TOCO,
see~\cite{groundexp}; see also~\cite{tegmark_web} for an up-to-date
compilation of the current results) have by now probed the
degree-scale anisotropy of the microwave sky. They seem to indicate
the presence of a high peak in the spectrum. Now, in contradiction
with these observations, topological defect models do not produce
generically much more power on the degree scale than on the largest
scales observed by the COBE satellite~\cite{cobe}. The reason is that,
although the defects evolve according to causal processes, thus
producing power on small angular scales on the last scattering
surface, one must also take into account their gravitational
interaction with the photons {\it after} last scattering. This
so-called Integrated Sachs-Wolfe effect (see
\EG{}~\cite{hu}), known to be negligible in most inflationary
scenarios, greatly increases in the case of defects the power on large
angular scales, thus contradicting the present observations (see
however \cite{vachas}, where the wiggly structure of strings is
properly taken into account, yielding more power on the degree
scale). Moreover, in most defect theories, the situation is also
worsened by the fact that scalar, vector and tensor modes all
contribute significantly to the overall CMB anisotropies, whereas only
the scalar contribution is expected to produce more power on small
angular scales than on larger ones.

It must however be stressed that, in all the numerical defect models
already mentionned, an important physical effect, to wit their decay
into gravitational radiation and/or elementary particles, has been
considered in most cases in a phenomenological way only (but see
however~\cite{damour}). Turok \ETAL{} and Durrer \ETAL{} have imposed
and checked that their global defect stress-energy tensor is
conserved. In cosmic string numerical simulations one has to deal with
the problem of loop production and evolution, which has been treated
in various ways. Turok \ETAL{} have introduced an extra radiation
fluid into which their cosmic string loops are supposed to decay,
Battye \ETAL{} have introduced an extra fluid with given constraints
on its stress-energy tensor, and Magueijo \ETAL, whose stress-energy
tensor is not conserved due to the fact that the smallest loops are
extracted out of the simulation, ``dump'' the string energy losses
into either extra-fluids with no anisotropic stress and various
equations of state, or in one or the other of the existing background
fluids, that is the baryons, photons, neutrinos or CDM. Finally,
Shellard \ETAL{} treat the loops as relativistic point masses, and
Pogosian \ETAL{} treat them as small segments.

The aim of this paper is first to show that the microphysics of
defects imposes that energy be released in the cosmic fluids and that
the predictions concerning the CMB anisotropies change drastically
when even a small fraction of the network energy is released directly
in the form of photons or baryons, and, to a lesser extent,
neutrinos. As for the matter power spectrum, we will see that it is
greatly affected if the network energy is released into photons,
neutrinos, or baryons.

Such a lack of generic predictive power for seeds-based models has
already been pointed out in another context by Durrer and
Sakellariadou~\cite{durmair}, who have used various ans\H{a}tze for
the stress-energy tensor components of scaling coherent seeds and have
found a wide variety of results for the positions and heights of the
induced CMB anisotropies Doppler peaks. One could argue that these
results were obtained by using very ad-hoc hypothesis for the seeds
correlators, but at least three other works have reached similar
conclusions by using more realistic
models. Perivolaropoulos~\cite{perivar} has pointed out that a correct
modelling of the gravitational interaction between cosmic strings
wiggles and electrons could induce more small scale anisotropies than
previously expected, and this analysis was confirmed by the more
precise numerical works of Pogosian and
Vachaspati~\cite{vachas}. Finally, the inclusion of decay products in
the specific model of Contaldi \ETAL{}~\cite{magueijo} proved to have
an important effect on the CMB anisotropies spectrum. Interestingly
enough, these two models which add some small-scale physics give a
better agreement with observations than those which do not include any
small-scale physics. We show here that such effects are not restricted
to any specific model but hold for a very large class of defects and
decay processes.

We shall first discuss the microphysics behind such effects, then show
how one can generically take them into account, and finally present
some results.

\section{The decay products of realistic string models}
\label{sec_decay}
 
Topological defects fall in various classes~\cite{td}.

Uncharged Goto-Nambu strings form loops which are usually supposed to
decay dominantly into gravitational
radiation~\cite{rad_Goldstone,allen96}, thus guaranteeing the scaling
behaviour of the network (that is the fact that the ratio of the
energy density contained in the network with the total energy density
is constant in time). This extra gravitational radiation added to the
background is observationally constrained by \EG{} the millisecond
pulsar timing measurements~\cite{maitia}.

Uncharged global strings, when seen from a distance, appear very much
like local ones, but with an energy per unit length renormalized to
include long-range interaction effects~\cite{global}. The overall
global string network is therefore expected to behave much like a
Goto-Nambu network, except for the small loops~\cite{renorm}. The main
difference between a local and a global string loop is that while the
former is supposed to radiate mostly in gravitational waves, the
latter looses energy mainly through massless Goldstone boson
radiation, a process known to be far more
efficient~\cite{renorm}. This new extra component added to the
background is only constrained by
nucleosynthesis~\cite{constraint_glob}.

The last potentially interesting class of strings in the context
considered here is that of Grand-Unified (GUT) superconducting cosmic
strings~\cite{witten} (only GUT strings can be relevant in large scale
structure formation and CMB fluctuations). Again, their network
evolution is almost the same as that of Goto-Nambu strings because
most of the evolution takes place when the currents flowing along the
strings are negligible~\cite{witt_evol}. This is however true only for
infinite strings or for those having very large radius of curvature.

When a superconducting loop decays, its energy can be very efficiently
released in the background {\it directly} in the form of
electromagnetic radiation~\cite{elec_rad}. The electromagnetic
radiation cannot propagate because of the surrounding plasma, the
influence of which cannot be neglected. The waves form shells where
energy is concentrated. These shells are the basis for the explosive
Ostriker, Thompson and Witten large scale structure formation
model~\cite{otw} and yield a distorsion in the microwave background by
modifying its spectrum, implying non zero values for the parameters
$\mu$ (chemical potential distortion, see \EG~\cite{hu_silk}) and $y$
(characterizing the Sunyaev-Zel'dovitch effect, see
\EG~\cite{sz}). This very stringent constraint, together with \EG{}
nucleosynthesis constraints~\cite{sigl} almost rules the scenario out,
leaving as the only possibility that the radiation can only be emitted
at much higher frequencies (to allow propagation in the plasma). This
requires that the strings can be considered current-free until they
have shrunk to a sufficiently small size, the wavelength of the
radiation being proportional to the emitting loop radius. As a result,
one can consider the network evolution as, again, that of a Goto-Nambu
network with the difference that part of the energy contained in the
loops can now be transferred {\it directly} into the photon fluid.

When such a loop shrinks however, the integrated current being
conserved, the energy it contains per unit length increases and its
effects become more and more important on the string dynamics. The
resulting loop distribution can accumulate to stationary states known
as vortons~\cite{vortons}. This overproduction of vortons breaks the
scaling behaviour of the network, thus leading to such a cosmological
catastrophe that one must assume that the vortons are sufficiently
unstable to decay in less than one Hubble time.

The new phenomenon one therefore needs to consider is the fate of the
small charged loops. As discussed above, they can decay into
gravitational radiation, or directly into photons. The third, largely
overlooked possibility when it comes to compute CMB anisotropies or
matter power spectrum, is that they might decay into their
constituents~\cite{xav4}, namely massive Higgs and gauge bosons and
the particles they couple to and that make the current, collectively
referred to as ``$X$''-particles. As these particles have masses
comparable to the grand unification scale, their decay products can
only be estimated by the relevant QCD extrapolations at high
energy~\cite{QCD}. The standard scenario is that the primary
$X$-particle will decay into a lepton (usually an electron) and a
quark which subsequently initiate a hadronic shower. Once the shower
has evolved, one ends up with roughly $3$\% nucleons~\cite{QCD}, and
the rest in $\pi$'s, which, because of their decays or interactions
with other background fluids, turn into neutrinos, photons and
electrons. These interactions have even been used to try and explain
the ultra high energy cosmic ray~\cite{UHECR} enigma by means of
topological defects~\cite{TD_UHECR}. Note that this picture does not
take into account the possibility that part of the decay product be a
stable particle, \IE{} a constituent of the dark matter. The Lightest
Super Particle is a possible candidate, for instance if SUSY is
demanded, as should be the case for any GUT model.

This analysis of small superconducting loops has led us to conclude
that they mostly decay (hence ensuring that their network scales) {\it
directly} into the constituents of the universe rather than into
extra-fluids such as gravitational radiation.

In fact one can also argue that {\it non} superconductiong strings,
because they intercommute, can also partly decay into photons,
baryons, neutrinos or dark matter, and not only into gravitational
radiation or Goldstone bosons for the global ones. Even global defects
can also decay~\cite{td}~: when the gradients are strong enough high
energy particles and gravitational radiation are produced. One might
expect however that the energy losses in the form of background fluids
be far more efficient for superconducting, rather than uncharged,
defects.

\section{A two-parameters decay model}
\label{sec_model}

When one solves the classical (Goto-Nambu) equations of motion for the
string network in a Friedmann-Robertson-Walker (FRW) background,
completed by a set of rules which fixes the intercommutation of long
strings into loops, in the stiff approximation~\cite{stiff} framework
(in which the defect network energy is a first order perturbation to
the FRW background), and if no extra physics is added to the problem,
the corresponding stress-energy tensor of the strings {\it must} be
covariantly conserved. If it is not, that only means that the
numerical integration is not precise enough. The same holds true for
global defects.

Now, as we have seen in \S\ref{sec_decay}, extra physics {\it must} be
added. Indeed intercommutation leads to the formation of more and more
small loops per Hubble volume which would prevent scaling to take
place and soon dominate the evolution of the universe, were they not
eliminated by turning into gravitational radiation and/or various
elementary particles. The string stress-energy tensor cannot then be
conserved. This is particularly clear in numerical simulations when
loops smaller than a given size are phenomenologically extracted by
hand from the network (as in \EG~\cite{magueijo}).

In the semi-analytic approach we use here (see below), we model active
seeds by a stress-energy tensor $\Theta^s_{\mu\nu}$ which, from the
start, embodies the scaling properties of the network~: first, deep in
the radiation-dominated or matter-dominated eras, its equal-time
correlators $\left<\Theta^s_{\mu\nu}({\mathbf x}, \eta)
\Theta^s_{\rho\sigma}({\mathbf x'}, \eta)\right>$ depend on $({\mathbf
x}-{\mathbf x'})/\eta$ only, second they vanish at small scales
($\eta$ is conformal time and $x^i$ are comoving
coordinates). Therefore this effective stress-energy tensor describes
in fact, not only the defects themselves but also their decay
products, as long as these decay products are themselves ``seeds'',
that is an active perturbation added to the background fluids. Now
among the possible decay products described in \S\ref{sec_decay}, only
the gravitational and Goldstone radiations enter in that
category. Hence, if the defects were scaling only via the production
of this ``extra'' radiation, the effective stress-energy tensor we
use, being the sum of the stress-energy tensor of the defects and that
of the extra radiation they produce, would be conserved. In contrast,
in numerical simulations where the defect stress-energy tensor is not
conserved, such decay products must be treated as an extra fluid with
given equation of state.

However, as we have seen, in the case of superconducting strings at
least, loops decay mainly through the emission of high energy
particles which soon turn into (mostly) neutrinos and photons, which
must be added to those already existing in the background. Therefore
we shall describe the defects by an effective stress energy-tensor
which will not be conserved in order to take into account the decay of
the loops into high energy particles, and we shall dump the string
energy losses not in an extra component but into the background
fluids.

In the semi-analytic approach initiated by Durrer and collaborators
\cite{durrer}, the defect network and the extra ``seeds'' they decay
into are modelled by a stress-energy tensor $\Theta_{\mu\nu}^s$ which
acts as a source for the linearized Einstein equations and induces
inhomogeneities in the cosmic fluids. The ten components of this
stress-energy tensor are then drastically constrained by a number of
physical requirements~\cite{turok4,dlu,udr}, which take into account
that the seeds~:
\begin{enumerate} 

  \item \label{enum1} are statistically homogeneous and isotropic,

  \item \label{enum2} are created at a phase transition in an up to
  then perfectly homogeneous and isotropic universe (hence obeys
  specific causality \cite{turok4} and matching \cite{dlu}
  conditions),

  \item \label{enum3} evolve deep in the radiation era and deep in the
  matter era in a way which is statistically independent of time
  (scaling requirement \cite{udr}).

\end{enumerate}

One can now suppose that the defect network is made of long strings of
comoving curvature radius $R_L > 1/\Hconf$ ($\Hconf$ is the comoving
Hubble parameter), which eventually interconnect so as to form
loops. When they become too small these loops decay at a certain rate
(depending on the details of the decay processes). The loss of energy
of the stress-energy tensor can therefore, in a rough approximation,
be modelled as
\begin{eqnarray}
\label{nc1}
D_\mu \Theta_s^{\mu\nu} & = & F^\nu, \\
\label{nc2}
\hat F^\nu = -\XM \hat\Theta_s^{0\nu}&\quad\mathrm{with}\quad&
\XM = \XH Y(k - \XK) 
\end{eqnarray}
where $D_\mu$ is a covariant derivative (we assume flat spatial
section), a hat denotes a Fourier transform, $Y$ is the Heavyside
function, $k \propto 1/R_L$ is the comoving wavenumber. $\XH$ is a
free function which determines the decay rate of the loops into the
decay products, and $\XK$ is the scale above which the loops begin to
decay. This $\hat F^\nu$ is then injected into the other fluid
perturbation equations with branching ratios $x_f$, so as to ensure
that the total stress-energy tensor is conserved (see appendix
\ref{detail} for the actual equations we solve). The
branching ratios $x_f$ are in principle calculable when the decay
microphysics is explicited.

\section{Results}

\subsection{CMB anisotropies}

We have computed numerically the CMB anisotropies and the matter power
spectrum such decaying seeds produce. We first summarize our findings
before going into discussing them. Dumping energy into CDM has
negligible effect on the CMB anisotropies spectrum. Dumping energy
into neutrinos has a relatively small influence on the spectrum, which
is modified by a few tens of percent. This can either increase or
decrease the amplitude of the spectrum [we have found that this
depends on the parameters of the decay term~(\ref{nc2}), as well as on
the ans\H{a}tze for the correlators~(\ref{ps},\ref{pis})]. On the
contrary, the spectrum is tremendously affected by injecting energy
into either photons or baryons, although the effect is much stronger
when one injects energy directly into photons, which can boost the
degree-scale amplitude of the spectrum by a factor as large as 50. The
precise amplitude of the boost depends of course on the underlying
defect model, but all those we have checked present the same
qualitative features.

A simple interpretation of these results is that energy, when injected
into either neutrinos or CDM, influences the photons perturbations
only through gravity, so that the influence of this energy injection
on the CMB anisotropies is small. On the opposite, by dumping energy
into baryons or photons, one directly affects the evolution of the
perturbed photons density and/or velocity since photons and baryons
are strongly coupled through Thomson diffusion before
recombination. The angular scale at which this effect is strongest is
given by the angular size of the ``decay scale'' (\IE{} the Hubble
radius when $\XK=\Hconf$) at the last scattering surface, that is
$\theta \simeq 1 \deg$.

\begin{figure}
\begin{center}
\begingroup%
  \makeatletter%
  \newcommand{\GNUPLOTspecial}{%
    \@sanitize\catcode`\%=14\relax\special}%
  \setlength{\unitlength}{0.12bp}%
\begin{picture}(3600,2160)(0,0)%
\special{psfile=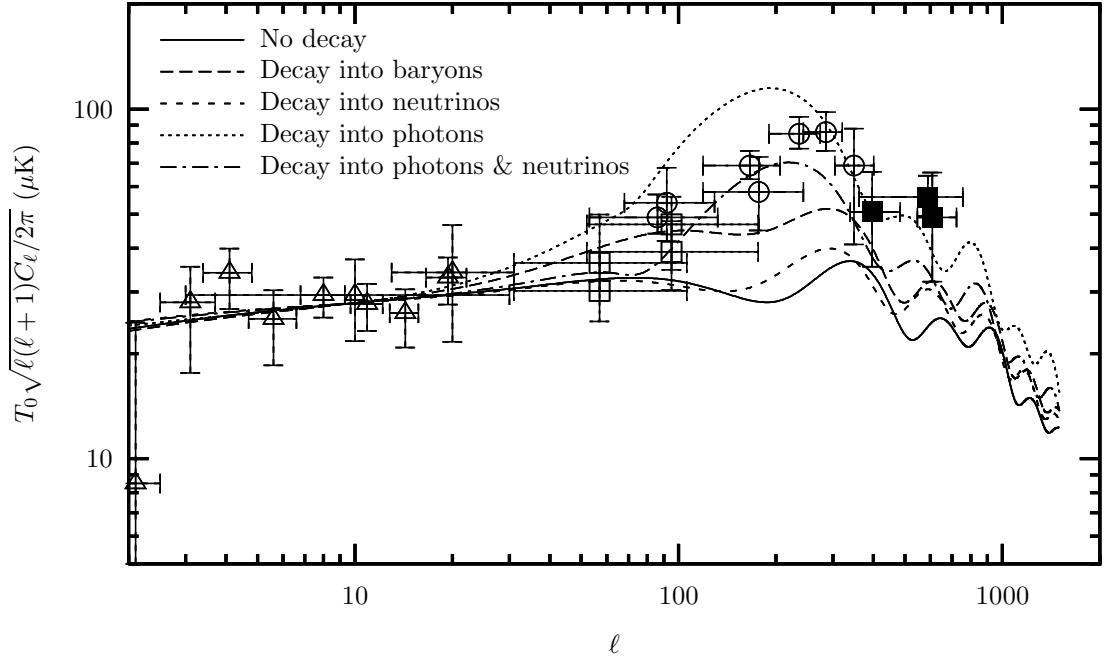 llx=0 lly=0 urx=600 ury=420 rwi=7200}
\put(813,1547){\makebox(0,0)[l]{Decay into photons \& neutrinos}}%
\put(813,1647){\makebox(0,0)[l]{Decay into photons}}%
\put(813,1747){\makebox(0,0)[l]{Decay into neutrinos}}%
\put(813,1847){\makebox(0,0)[l]{Decay into baryons}}%
\put(813,1947){\makebox(0,0)[l]{No decay}}%
\put(1925,50){\makebox(0,0){$\ell$}}%
\put(100,1180){%
\special{ps: gsave currentpoint currentpoint translate
270 rotate neg exch neg exch translate}%
\makebox(0,0)[b]{\shortstack{$T_0 \sqrt{\ell(\ell+1) C_\ell / 2\pi}$ ($\mu$K)}}%
\special{ps: currentpoint grestore moveto}%
}%
\put(3144,200){\makebox(0,0){1000}}%
\put(2127,200){\makebox(0,0){100}}%
\put(1111,200){\makebox(0,0){10}}%
\put(350,1729){\makebox(0,0)[r]{100}}%
\put(350,631){\makebox(0,0)[r]{10}}%
\end{picture}%
\endgroup
 
\end{center}
\caption{CMB anisotropies in a model where the defect stress-energy
tensor is not conserved, with $\XK=\XH=\Hconf$. The solid line
represents the case with conserved stress-energy tensor, as well as
the case where one dumps energy into CDM, which are almost
identical. The short-dashed and long-dashed lines represent the cases
where one dumps energy into neutrinos and baryons respectively. The
highest (dotted) line represents the case where energy is injected
into photons. The dot-dashed line (which fits the data points best)
shows that by tuning by hand the branching ratios (here, $x_\nu =
x_\gamma = 0.5$ and $x_c = x_b = 0$) it is possible to be in much
better agreement with the data points. Note that the precise values of
the best-fit branching ratios depend on the model one considers.}
\label{fig1}
\end{figure}

For example in \FIG{\ref{fig1}}, we consider the scalar contribution
to the CMB anisotropies of a specific coherent model where we have
chosen the equal time correlators for the pressure and the anisotropic
stress [see~\EQNS{\ref{ps},\ref{pis}}] in such a way that they respect
the standard causality and scaling requirements (see appendix for more
details). We have also chosen the relative amplitudes of the
correlators so that, when the stress-energy tensor is conserved, we
``mimic'' the situation of more realistic incoherent numerical defect
models (\IE{} the CMB anisotropies spectrum lacks power on the
degree-scale with respect to observations; the coherent model
presented here exhibits acoustic oscillations which are expected to be
smoothed out by decoherence, see~\cite{turok}). The two parameters of
the decay model, $\XK$ and $\XH$ [see \EQ{\ref{nc2}}] have both been
chosen equal to $\Hconf$. We have also studied many other models such
as the ``pressure model''~\cite{cheung} [where ones sets the
anisotropic stress $\hat\Pi^s$ to 0, see \EQNS{\ref{ps},\ref{pis}}],
or the ``anisotropic stress model'' (where the pressure $\hat P^s$ is
set to 0, see \EG~\cite{hu_whi}), which all yield similar qualitative
results. In all what follows, the CMB data points are taken
from~\cite{tegmark_web} and the matter power spectrum data points are
those of the APM catalog~\cite{apm} and of Dekel~\ETAL~\cite{dekel}.

\begin{figure}
\begin{center}
\begingroup%
  \makeatletter%
  \newcommand{\GNUPLOTspecial}{%
    \@sanitize\catcode`\%=14\relax\special}%
  \setlength{\unitlength}{0.12bp}%
\begin{picture}(3600,2160)(0,0)%
\special{psfile=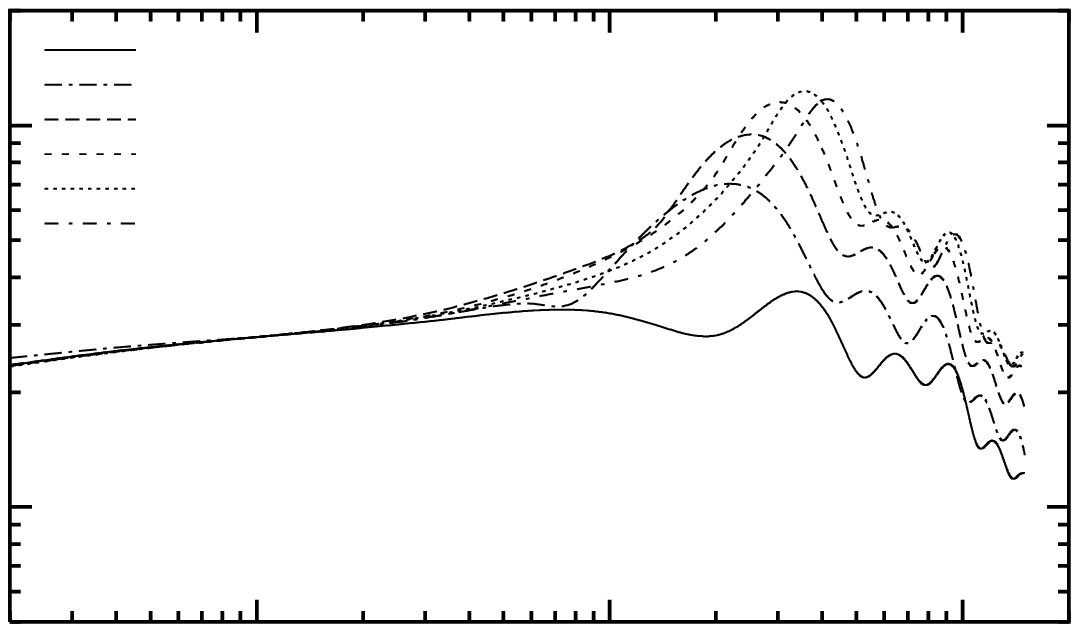 llx=0 lly=0 urx=600 ury=420 rwi=7200}
\put(813,1447){\makebox(0,0)[l]{id, $k_c = 5{\mathcal H}$}}%
\put(813,1547){\makebox(0,0)[l]{id, $k_c = 4{\mathcal H}$}}%
\put(813,1647){\makebox(0,0)[l]{id, $k_c = 3{\mathcal H}$}}%
\put(813,1747){\makebox(0,0)[l]{id, $k_c = 2{\mathcal H}$}}%
\put(813,1847){\makebox(0,0)[l]{Decay into photons \& neutrinos, $k_c ={\mathcal H}$}}%
\put(813,1947){\makebox(0,0)[l]{No decay}}%
\put(1925,50){\makebox(0,0){$\ell$}}%
\put(100,1180){%
\special{ps: gsave currentpoint currentpoint translate
270 rotate neg exch neg exch translate}%
\makebox(0,0)[b]{\shortstack{$T_0 \sqrt{\ell(\ell+1) C_\ell / 2\pi}$ ($\mu$K)}}%
\special{ps: currentpoint grestore moveto}%
}%
\put(3144,200){\makebox(0,0){1000}}%
\put(2127,200){\makebox(0,0){100}}%
\put(1111,200){\makebox(0,0){10}}%
\put(350,1729){\makebox(0,0)[r]{100}}%
\put(350,631){\makebox(0,0)[r]{10}}%
\end{picture}%
\endgroup
 
\end{center}
\caption{Influence of the decay scale $\XK$ on the CMB
anisotropies. We use the ``best fit'' model of \FIG{\ref{fig1}}, with
50\% of the energy relased into photons and another 50\% into
neutrinos, and we vary the parameter $\XK$ from $\Hconf$ to
$5\Hconf$. The solid line represents as in \FIG{\ref{fig1}} the case
where the seed stress-energy tensor is conserved, and the case
$\XK=\XH=\Hconf$ corresponds to the dot-dashed curve, also plotted in
\FIG{\ref{fig1}}.}
\label{fig2}
\end{figure}

\begin{figure}
\begin{center}
\begingroup%
  \makeatletter%
  \newcommand{\GNUPLOTspecial}{%
    \@sanitize\catcode`\%=14\relax\special}%
  \setlength{\unitlength}{0.12bp}%
\begin{picture}(3600,2160)(0,0)%
\special{psfile=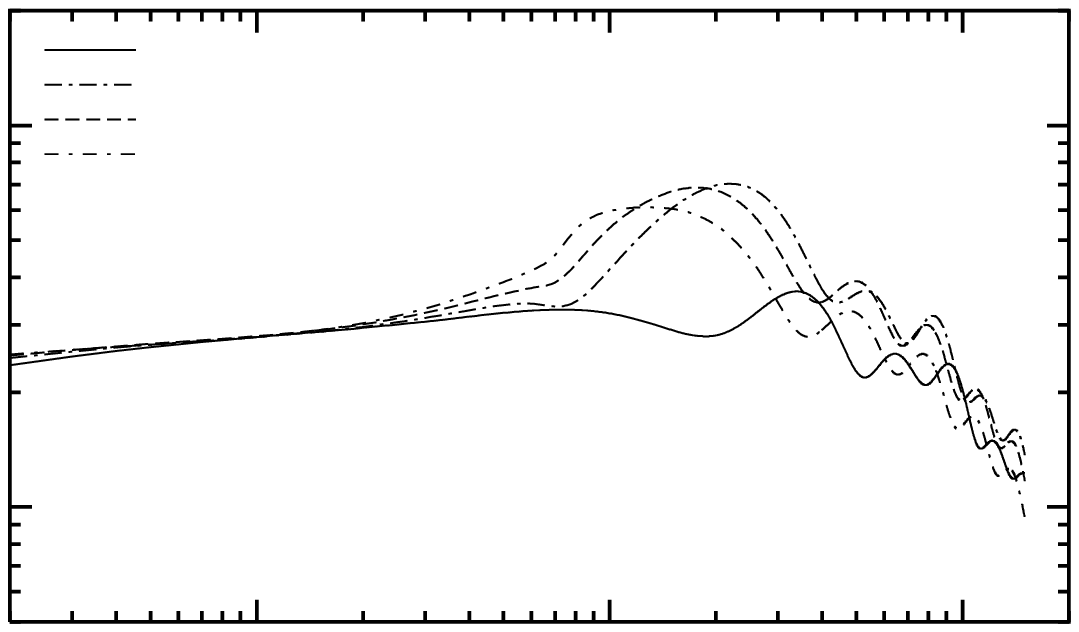 llx=0 lly=0 urx=600 ury=420 rwi=7200}
\put(813,1647){\makebox(0,0)[l]{id, $\Gamma = 5 {\mathcal H}$}}%
\put(813,1747){\makebox(0,0)[l]{id, $\Gamma = 2 {\mathcal H}$}}%
\put(813,1847){\makebox(0,0)[l]{Decay into photons \& neutrinos, $\Gamma = {\mathcal H}$}}%
\put(813,1947){\makebox(0,0)[l]{No decay}}%
\put(1925,50){\makebox(0,0){$\ell$}}%
\put(100,1180){%
\special{ps: gsave currentpoint currentpoint translate
270 rotate neg exch neg exch translate}%
\makebox(0,0)[b]{\shortstack{$T_0 \sqrt{\ell(\ell+1) C_\ell / 2\pi}$ ($\mu$K)}}%
\special{ps: currentpoint grestore moveto}%
}%
\put(3144,200){\makebox(0,0){1000}}%
\put(2127,200){\makebox(0,0){100}}%
\put(1111,200){\makebox(0,0){10}}%
\put(350,1729){\makebox(0,0)[r]{100}}%
\put(350,631){\makebox(0,0)[r]{10}}%
\end{picture}%
\endgroup
 
\end{center}
\caption{Influence of the decay rate $\XH$ on the CMB anisotropies. We
use the ``best-fit'' model of \FIG{\ref{fig1}}, with 50\% of the
energy relased into photons and another 50\% into neutrinos, and we
vary the parameter $\XH$ from $\Hconf$ to $5\Hconf$. The solid line
represents as in \FIG{\ref{fig1}} the case where the seed
stress-energy tensor is conserved, and the case $\XK=\XH=\Hconf$
corresponds to the dot-dashed curve, also plotted in
\FIG{\ref{fig1}}.}
\label{fig3}
\end{figure}

In \FIGS{\ref{fig2} and \ref{fig3}} we study the influence of the
parameters $\XK$ and $\XH$ for the model of \FIG{\ref{fig1}} where
energy is equally released into photons or neutrinos. By increasing
$\XK$ (\FIG{\ref{fig2}}), which gives essentially the wavenumber scale
at which energy is injected, we get that, as expected, the spectrum is
enhanced on smaller angular scales, \IE{} the bump is shifted to the
right. On the contrary, by increasing $\XH$ (\FIG{\ref{fig3}}), namely
the defects decay rate, energy is transferred more rapidly into the
background fluids, and therefore at larger wavelengths. The spectrum
is thus enhanced on larger angular scales, so that the bump is shifted
to the left.

\subsection{Matter power spectrum}

\begin{figure}
\begin{center}
\begingroup%
  \makeatletter%
  \newcommand{\GNUPLOTspecial}{%
    \@sanitize\catcode`\%=14\relax\special}%
  \setlength{\unitlength}{0.12bp}%
\begin{picture}(3600,2160)(0,0)%
\special{psfile=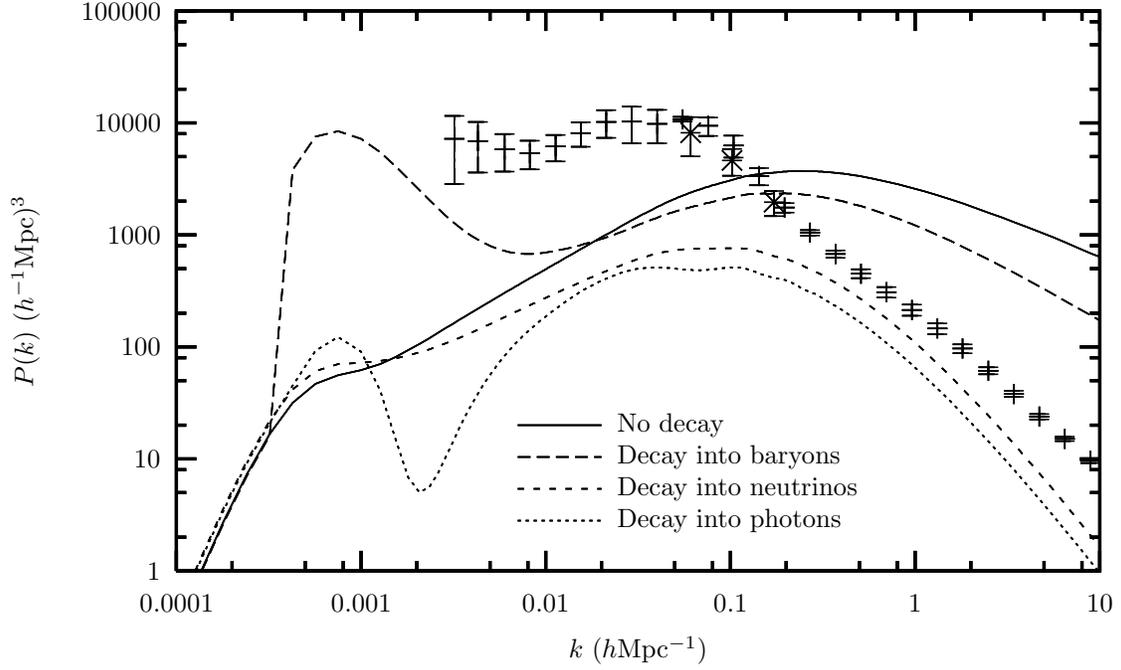 llx=0 lly=0 urx=600 ury=420 rwi=7200}
\put(1935,458){\makebox(0,0)[l]{Decay into photons}}%
\put(1935,558){\makebox(0,0)[l]{Decay into neutrinos}}%
\put(1935,658){\makebox(0,0)[l]{Decay into baryons}}%
\put(1935,758){\makebox(0,0)[l]{No decay}}%
\put(2000,50){\makebox(0,0){$k$ ($h$Mpc$^{-1}$)}}%
\put(100,1180){%
\special{ps: gsave currentpoint currentpoint translate
270 rotate neg exch neg exch translate}%
\makebox(0,0)[b]{\shortstack{$P(k)$ ($h^{-1}$Mpc)$^3$}}%
\special{ps: currentpoint grestore moveto}%
}%
\put(3450,200){\makebox(0,0){10}}%
\put(2870,200){\makebox(0,0){1}}%
\put(2290,200){\makebox(0,0){0.1}}%
\put(1710,200){\makebox(0,0){0.01}}%
\put(1130,200){\makebox(0,0){0.001}}%
\put(550,200){\makebox(0,0){0.0001}}%
\put(500,2060){\makebox(0,0)[r]{100000}}%
\put(500,1708){\makebox(0,0)[r]{10000}}%
\put(500,1356){\makebox(0,0)[r]{1000}}%
\put(500,1004){\makebox(0,0)[r]{100}}%
\put(500,652){\makebox(0,0)[r]{10}}%
\put(500,300){\makebox(0,0)[r]{1}}%
\end{picture}%
\endgroup
 
\end{center}
\caption{Matter power spectrum in the same model as in
\FIG{\ref{fig1}}, with $\XK=\XH=\Hconf$. The solid line represents the
case where the seed stress-energy tensor is conserved. Dumping energy
into neutrinos (short-dashed line) or photons (dotted line) has almost
the same effect, whereas dumping energy into CDM (solid line) has
again negligible effect. Dumping energy into baryons (long-dashed
line) boosts the spectrum on larger scales, whereas it damps it on
smaller scales. Note that the sharp bump around $k=10^{-3}$
$h$Mpc$^{-1}$ shows the scale at which one injects energy (here, the
Hubble radius since $\XK=\Hconf$).}
\label{fig4}
\end{figure}

Dumping energy into CDM has also negligible effect on the (baryonic)
matter power spectrum. On the contrary, the spectrum is strongly
affected by injecting energy into either photons or neutrinos. It has
the consequence of reducing the excess of energy on small scales,
because of the free streaming of these relativistic particles (see
\EG{}\cite{padm}). Finally, injecting energy into baryons gives an
intermediate result. \FFIG{\ref{fig4}} summarizes these results by
showing the matter power spectrum corresponding to the model of
\FIG{\ref{fig1}}.

\subsection{Observational constraints}
\label{sec_obs}

In all these numerical analysis, we have assumed that the radiation
emitted by the high energy decay product is immediately thermalized as
soon as it is produced. However this is only an approximation. Three
unwanted effects can be caused by this radiation. First, the radiation
emitted before the last scattering surface may not have had the time
to thermalize, thus leading to a distortion of the CMB black-body
spectrum; this would be the case if too much energy is injected
between $z\simeq 10^6$ and $z\simeq 10^3$. However we inject a small
amount of energy compared to that already contained in the CMB
(roughly $10^{-6}$ per Hubble time since one is in the
radiation-dominated era), so that (considering the current
observational bounds on the $\mu$ and $y$ parameters) one can assume
it had time to thermalize, see~\cite{hu_silk}. Second, the radiation
emitted after the last scattering surface will not be thermalized, and
therefore could produce an important $\gamma$-ray
background~\cite{sigl}. Third, the high-energy particles could
photo-dissociate the $^4$He nuclei, thus producing lighter nuclei such
as $^3$He and D, which in turn can produce too much
$^6$Li~\cite{jedam}.

\begin{figure}
\begin{center}
\centerline{\psfig{file=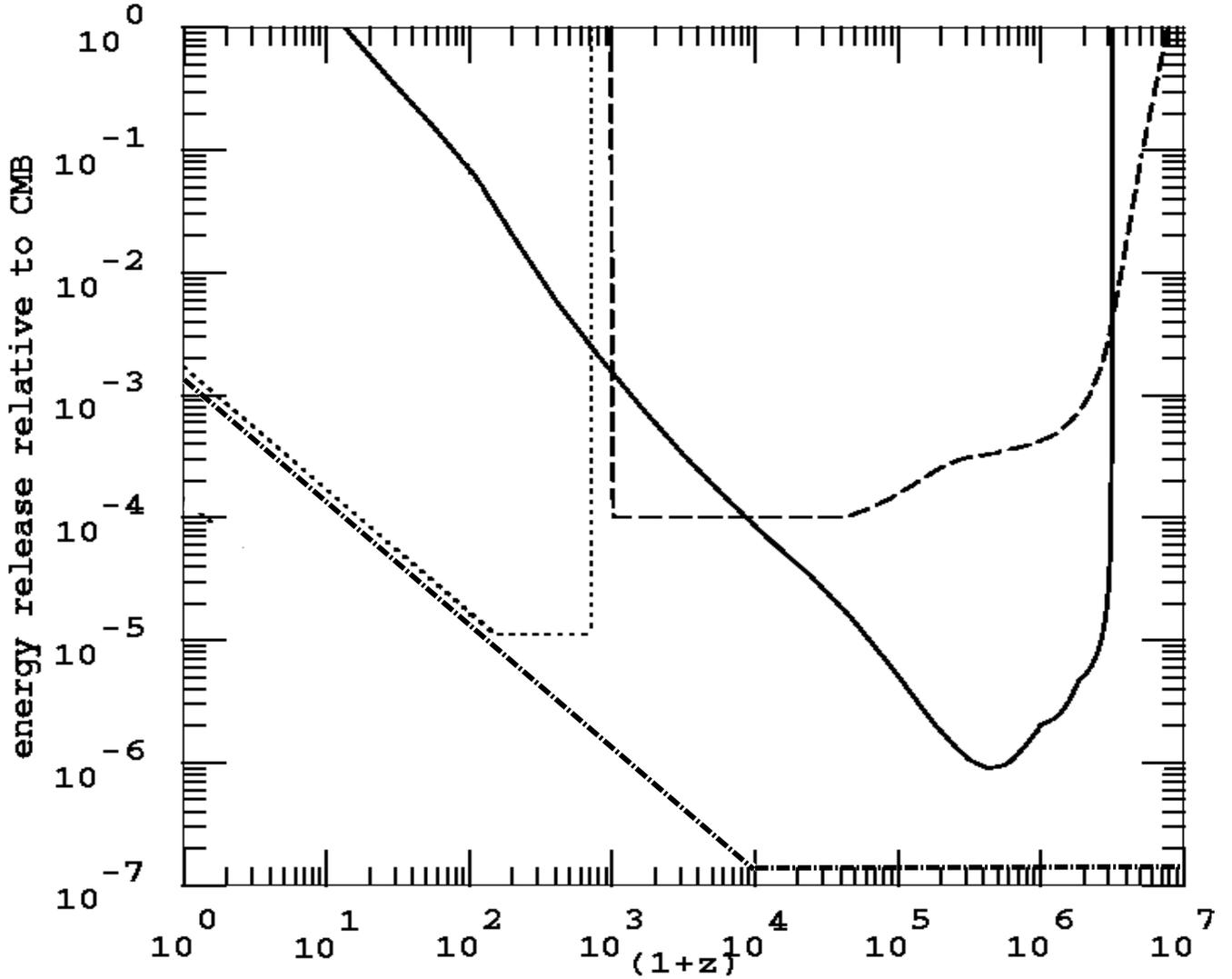,width=7in,angle=0}}
\end{center}
\caption{Maximal energy release in units of the CMB energy density
allowed by the constraints from the observed diffuse $\gamma-$ray
background at $5\,$GeV (dotted curve), CMB distortions (dashed curve),
and $^4$He photo-disintegration (solid curve) as a function of
redshift $z$ (taken from Sigl \ETAL). These bounds apply for
instantaneous energy release at the specified redshift epoch. The
lowest (dot-dashed) curve shows the energy release in our model
assuming that $\simeq$10\% of the energy is released into photons
(energy releases into the other fluid do not add to these
constraints). Higher energy release into photons might violate the
$\gamma$-ray background measurements. Note that the strongest
constraint arise today from the diffuse $\gamma$-ray background, but
better measurements of the CMB spectrum could give an even stronger
constraint.}
\label{fig5}
\end{figure}

These three effects are already constrained by the observations of the
CMB spectrum, the diffuse $\gamma$-ray background and the light
elements abundances, but our scenario still happens to be tenable, as
shown on \FIG{\ref{fig5}} (taken from~\cite{sigl}). In the opposite
case, this would disproof most scenario of structure formation seeded
by topological defects which dominantly decay into photons.

\section{Conclusion} 
\label{sec_conclusion}

We have included some microphysics, up-to-now largely overlooked, in
the description of topological defects. This microphysics deals with
the decay of defects, and notably superconducting cosmic strings, into
background fluids, rather than gravitational radiation as usually
assumed. This decay has important observational consequences, which
may (depending, of course, on the exact interaction one considers) put
the defect models in a better position when confronted to the current
observational data.

One could of course argue that the simple model we have considered
here (coherent seeds, crude interaction term) is too naive, however
our purpose was mainly to illustrate the consequences of this idea
rather than to study a more specific realistic model, which we keep
for later work. Since the main consequences of dumping energy into the
background fluids do not seem to depend too much on the details of the
equal time correlators we have considered, but rather on the details
of the interactions, we strongly advocate for a more careful study of
the branching ratios.

Let us recall that even in inflationary models the inclusion of
microphysics and interaction between fluids is absolutely crucial in
order to make accurate predictions. For example, if one neglects these
by not solving the exact Boltzmann equation for the photons and/or the
neutrinos, or by not solving the accurate kinetic recombination
equations, one finds an large excess of power at small angular scales
(see \EG{} \FIG{4} of~\cite{hu}).

It is therefore clear that predictions of topological defect models
concerning structure formation not only require today's
state-of-the-art heavy detailed numerical simulations, but also a
rigorous description of their non gravitational interactions before
reliable conclusions can be drawn.

\acknowledgements

We are happy to thank G\H{u}nter Sigl and Thibault Damour for
enlightening discussions.

~\\
\texttt{Alain.Riazuelo@obspm.fr}~\\
\texttt{Nathalie.Deruelle@obspm.fr}~\\
\texttt{Patrick.Peter@obspm.fr}

\appendix

\section{CMB anisotropies calculations}
\label{detail}

We assume standard cosmological parameters~: flat Universe without
cosmological constant, baryon density and Hubble parameter such that
$\Omega_b = 0.05$, $h = 0.5$, three massless neutrinos and standard
recombination.

The stress-energy tensor of the seeds can be decomposed as a sum of
scalar, vector and tensor~\cite{bardeen} random fields. The scalar
part, the only one we shall consider here, can be written in terms of
four random fields $\hat\rho^s$, $\hat v^s$, $\hat P^s$ and
$\hat\pi^s$, as~:
\begin{eqnarray}
\label{pert_deb}
K \hat\Theta_{00}^s & \equiv & \hat\rho^s ,\\
K \hat\Theta_{0i}^s & \equiv & -i k_i \hat v^s ,\\
K \hat\Theta_{ij}^s & \equiv &
   \hat P^s \delta_{ij}
 + \left(\frac{1}{3}k^2\delta_{ij}-k_i k_j\right)\hat\Pi^s .
\end{eqnarray}
($K$ is Einstein's constant.) To compute the CMB anisotropies, we need
the two-point correlators of the stress-energy tensor. The homogeneity
of the distribution imposes~:
\begin{equation}
\left<\hat\Theta^s_{\mu\nu}({\mathbf k},\eta)
      \hat\Theta^s_{\rho\sigma}({\mathbf k'},\eta')\right>
 = \delta({\mathbf k} - {\mathbf k'}) 
   \hat C_{\mu\nu\rho\sigma}({\mathbf k}, \eta, \eta'),
\end{equation}
where the ${\mathbf k}$-dependence of $\hat C$ is fixed by the
requirement that the distribution is isotropic. The correlators can be
decomposed as sums of ``coherent eigenmodes''~\cite{turok}~:
\begin{equation}
\hat C_{\mu\nu\rho\sigma}({\mathbf k}, \eta, \eta') = \sum_{(i)}
\lambda^{(i)} \hat c_{\mu\nu}^{(i)}({\mathbf k}, \eta) 
              \hat c_{\rho\sigma}^{(i)}({\mathbf k}, \eta'),
\end{equation}
where the $\hat c_{\mu\nu}^{(i)}({\mathbf k}, \eta)$ are the
correlators of coherent sources. The four random fields that describe
coherent sources are all proportional to the same normalised random
variable $e({\mathbf k})$ and hence are described by four random
fields. Two of these functions are constrained by
\EQNS{\ref{nc1}-\ref{nc2}}, and the two other must be either imposed
by hand or determined using some more detailed modelling. We take
simple ans\H{a}tze for the pressure $\hat P^s$ and the scalar
anisotropic stress $\hat\Pi^s$, in practice (up to a normalization
constant)~:
\begin{eqnarray}
\label{ps}
\hat P^s & = &
 \eta^{-\frac{1}{2}} \exp(-k^2 \eta^2) e({\mathbf k}) ,\\
\label{pis}
\hat \Pi^s & = &
 - 4 \eta^{-\frac{1}{2}} (k^2\eta^2) \exp(-k^2\eta^2) e({\mathbf k}) .
\end{eqnarray}
This (rather arbitrary) choice of correlators satisfies the
requirements imposed by causality, scaling, \ETC~\cite{udr}, and is
tailored in order to obtain a CMB anisotropies spectrum which lacks
power on the degree scale. Of course, the precise shape of these
functions can in principle be obtained by numerical simulations, and
the CMB anisotropies are known to depend very much on the shape of
these free functions~\cite{durmair}, but, as already stressed, we have
carefully checked that the effect we are interested in does {\it not}
qualitatively depend on them.

We solve (in the flat-slicing gauge) the well-known (see
\EG~\cite{dlu,udr,pertgen}) linearised Einstein equations which couple
the seeds network to the cosmic fluid inhomogeneities, corrected by
the decay term (\ref{nc1}-\ref{nc2}). They read~:
\begin{eqnarray}
\hat{\dot\rho}^s
 = - \Hconf \hat \rho^s -3 \Hconf\hat P^s 
   + k^2 \hat v^s
   -\XM \hat \rho^s
&\quad,\quad&
\hat{\dot v}^s
 = - 2 \Hconf \hat v^s 
   - \hat P^s + \frac{2}{3}\hat\Pi^s
   - \XM \hat v^s , \\
\hat{\dot\delta}_\gamma
 =   \frac{4}{3}k^2 \hat{v}_\gamma
   + x_\gamma \XM \frac{\hat \rho^s}{3\Hconf^2 \Omega_\gamma} 
&\quad,\quad&
\hat{\dot v}_\gamma
 = - \frac{1}{4} \hat{\delta}_\gamma
   + \frac{1}{6} k^2 \hat{\Pi}_\gamma
   - \hat\pbp-\hat\pbq
   - \dot\kappa\left(\hat{v}_\gamma-\hat{v}_b\right)
   + x_\gamma \XM \frac{\hat v^s}{4\Hconf^2 \Omega_\gamma} , \\
\label{pert_n}
\hat{\dot\delta}_\nu
 =   \frac{4}{3}k^2 \hat{v}_\nu
   + x_\nu \XM \frac{\hat \rho^s}{3\Hconf^2 \Omega_\nu} 
  &\quad,\quad&
\hat{\dot v}_\nu
 = - \frac{1}{4} \hat{\delta}_\nu
   + \frac{1}{6} k^2 \hat{\Pi}_\nu
   - \hat\pbp-\hat\pbq
   + x_\nu \XM \frac{\hat v^s}{4\Hconf^2 \Omega_\nu} , \\
\label{pert_c}
\hat{\dot\delta}_c
 =   k^2 \hat{v}_c
   + x_c \XM \frac{\hat \rho^s}{3\Hconf^2 \Omega_c} 
  &\quad,\quad&
\hat{\dot v}_c
 = - \Hconf \hat{v}_c
   - \hat\pbq
   + x_c \XM \frac{\hat v^s}{3\Hconf^2 \Omega_c} , \\
\label{pert_b}
\hat{\dot\delta}_b
 =   k^2 \hat{v}_b
   + x_b \XM \frac{\hat \rho^s}{3\Hconf^2 \Omega_b} 
  &\quad,\quad&
\hat{\dot v}_b
 = - \Hconf\hat{v}_b
   - \hat\pbq
   - \frac{4}{3}\frac{\rho_\gamma}{\rho_b}
     \dot\kappa\left(\hat{v}_b-\hat{v}_\gamma\right)
   + x_b \XM \frac{\hat v^s}{3\Hconf^2 \Omega_b},
\end{eqnarray}
where $\Omega_f$, $\delta_f$, $v_f$, $\Pi_f$ are respectively the
density parameter, the density contrast, the velocity and the
anisotropic stress pertubations for the fluid $f$, the subscripts $b$,
$c$, $\nu$, $\gamma$ mean respectively the baryonic, Cold Dark Matter,
neutrino and photon fluid, $\pbp$ and $\pbq$ are the two Bardeen
gravitational potentials, $\dot\kappa$ is the photons differential
opacity~\cite{peebles}, and a dot denotes a derivation with respect to
the conformal time. The terms in $\dot\mu$ account for the seeds
decay, and the branching ratios $x_f$ are constants such that~:
\begin{equation}
\sum_f x_f = 1.
\end{equation}
In addition, the two equations for the Bardeen potentials are~:
\begin{eqnarray}
2 \left[-k^2 + 3(\dot\Hconf-\Hconf^2)\right]\hat\pbp & = & 
3\Hconf^2\sum_f{\Omega_f\left[   \hat{\delta}_f
                              - 3(1+\omega_f)\hat{v}_f
                        \right]} + \hat\rho^s - 3 \Hconf\hat v^s, \\
\hat\pbq & = & \hat\pbp
- 3 \Hconf^2\sum_f{\Omega_f\omega_f\hat{\Pi}_f}-\hat\Pi^s.
\end{eqnarray}
Finally, the photons and neutrinos fluids obey a Boltzmann equation
which reads (for $\ell \geq 2$, see \EG~\cite{ma_ber} or~\cite{hu_whi}
for more details)~:
\begin{eqnarray}
\dot{\hat\Delta_\gamma^\ell}
 & = &   \frac{k}{2\ell+1}\left(  \ell\hat\Delta_\gamma^{\ell-1}
                                - (\ell+1)\hat\Delta_\gamma^{\ell+1}\right)
       - \dot\kappa \left(
         \hat\Delta_\gamma^\ell - \delta_{\ell,2} \hat P_\gamma
                    \right), \\
\dot{\hat E_\gamma^\ell}
 & = &   \frac{k}{2\ell+1}
         \left(  \ell\sqrt{1-4/\ell^2}\hat E_\gamma^{\ell-1}
               - (\ell+1)\sqrt{1-4/(\ell+1)^2}\hat E_\gamma^{\ell+1}\right)
       - \dot\kappa \left(
         \hat E_\gamma^\ell + \sqrt{6}\hat P_\gamma \delta_{\ell,2}
                    \right), \\
\hat P_\gamma & = & \frac{1}{10}\left(  \hat\Delta_\gamma^2
                                      - \sqrt{6}\hat E_\gamma^2\right), \\
\dot{\hat\Delta_\nu^\ell}
 & = &   \frac{k}{2\ell+1}\left(  \ell\hat\Delta_\nu^{\ell-1}
                                - (\ell+1)\hat\Delta_\nu^{\ell+1}\right),
\end{eqnarray}
with $\Delta_f^\ell$ and $E_f^\ell$ being respectively the $\ell$-th
moment of the temperature and electric-type polarisation distribution
functions of the species $f$, and where $P_f$ is the coupling between
temperature and polarization. The lowest multipoles of the
distribution function are related to the perturbed stress-energy
tensor components by~:
\begin{eqnarray}
\hat \delta_f & = & 4 \hat \Delta_f^0, \\
\hat v_f & = & -\frac{3}{k} \hat \Delta_f^1, \\
\label{pert_fin}
\hat \Pi_f & = & \frac{12}{k^2} \hat \Delta_f^2
\end{eqnarray}
\EEQNS{\ref{pert_deb}-\ref{pert_fin}} are solved numerically using a
Boltzmann code developped by one of us (A.R.) using standard initial
conditions~\cite{dlu}. The CMB anisotropies at present conformal time
$\eta_0$ are then calculated by the line-of-sight integration
method~\cite{cmbcode}~:
\begin{eqnarray}
\hat \Delta^\ell_\gamma(k, \eta_0) & = & \nonumber
\int_0^{\eta_0} \dot\kappa e^{-\kappa}
                \left( \frac{1}{4}\hat \delta_\gamma
                      +\hat \pbp +\hat \pbq\right)
                j_\ell(k(\eta_0-\eta)) \ddd \eta \\ \nonumber & + &
\int_0^{\eta_0} \dot\kappa e^{-\kappa}
                \left(- k \hat v_b\right)
                j'_\ell(k(\eta_0-\eta)) \ddd \eta \\ \nonumber & + &
\int_0^{\eta_0} e^{-\kappa}
                \left(\dot{\hat\pbp} + \dot{\hat\pbq}\right)
                j_\ell(k(\eta_0-\eta)) \ddd \eta \\            & + &
\int_0^{\eta_0} \dot\kappa e^{-\kappa} \frac{5}{2} \hat P_\gamma
                \left( 3 j''_\ell(k(\eta_0-\eta))
                      +  j_\ell(k(\eta_0-\eta))
                \right) \ddd \eta.
\label{sight}
\end{eqnarray}
In the line-of-sight integration formula~(\ref{sight}), we have
omitted the terms arising from the energy injection due to the seeds
decay into photons. These terms are proportional to
$e^{-\kappa}\dot\mu$ and therefore are generated between the last
scattering surface and today. The photons emitted because of this
decay have therefore very high energy and do not have time to
thermalize. Hence, they do not participate to the microwave
background, but rather to the diffuse $\gamma$-ray background, which
might put some interesting constraints on these models as discussed in
\S\ref{sec_obs}.

Finally, the scalar part of the 2-point correlator of the CMB
anisotropies is decomposed as~:
\begin{equation}
\left< \frac{\delta T}{T}({\mathbf n}_1)
       \frac{\delta T}{T}({\mathbf n}_2)\right>_
        {{\mathbf n}_1 . {\mathbf n}_2 = \cos(\theta)}
 = \frac{1}{4\pi} \sum_\ell (2\ell+1) C_\ell^S P_\ell(\cos \theta),
\end{equation}
where $P_\ell$ is the Legendre polynomial of order $\ell$, and the
coefficients $C_\ell^S$ are deduced from the photon distribution
multipoles by~:
\begin{equation}
C_\ell^S = 
 \frac{2}{\pi} \int |\hat\Delta^\ell_\gamma(k, \eta_0)|^2  k^2 \ddd k .
\end{equation}

\figure

\vfill\eject

\vfill\eject

\vfill\eject

\vfill\eject

\end{document}